\documentclass[twocolumn,preprintnumbers]{revtex4}

\usepackage{graphicx}
\usepackage{epsf}
\usepackage{amsmath,amssymb}
\usepackage{color}
\newcommand{\bvec}{\boldsymbol}

\usepackage{ulem}

\begin{document}
\preprint{KUNS-2754, NITEP 11}
\title{$\alpha$ scattering cross sections on $^{12}$C with microscopic coupled-channel calculation}

\author{Yoshiko Kanada-En'yo}
\affiliation{Department of Physics, Kyoto University, Kyoto 606-8502, Japan}
\author{Kazuyuki Ogata} 
\affiliation{Research Center for Nuclear Physics (RCNP), Osaka University,
  Ibaraki 567-0047, Japan}
\affiliation{Department of Physics, Osaka City University, Osaka 558-8585,
  Japan}
\affiliation{
Nambu Yoichiro Institute of Theoretical and Experimental Physics (NITEP),
   Osaka City University, Osaka 558-8585, Japan}
\begin{abstract}
$\alpha$ elastic and inelastic scattering on $^{12}$C is investigated with the
coupled-channel calculation using microscopic $\alpha$-$^{12}$C potentials, which are
derived by folding the Melbourne $g$-matrix $NN$ interaction with the matter and transition densities of $^{12}$C.
These densities are obtained by a microscopic structure model
of the antisymmetrized molecular dynamics
combined with and without the $3\alpha$ generator coordinate method.
The calculation reproduces satisfactorily well the observed elastic and inelastic cross sections
at incident energies of $E_\alpha=130$~MeV, 172.5~MeV, 240~MeV, and 386~MeV with no adjustable parameter.
Isoscalar monopole and dipole excitations
to the $0^+_2$, $0^+_3$, and
$1^-_1$ states in the $\alpha$ scattering are discussed.
\end{abstract}
\maketitle

\section{Introduction}

Cluster structure is one of the essential aspects of nuclear systems.
A variety of well developed cluster structures have been discovered
in excited states of stable light nuclei and also unstable nuclei.
In the past two decades, new types of multi-$\alpha$ cluster states have been theoretically
suggested in light $Z=N$ nuclei,
and experimental searching for new cluster states has been intensively
performed (see Refs.~\cite{Horiuchi:2012,Freer:2014qoa,Funaki:2015uya,Freer:2017gip} and references therein).

In the study of the nuclear clustering, $3\alpha$ cluster states in $^{12}$C have been
attracting a great interest for a long time \cite{Fujiwara80,Freer:2014qoa,Freer:2017gip}.
$3\alpha$-cluster models
suggested various cluster states near and above the $3\alpha$ threshold energy
\cite{Funaki:2015uya,kamimura-RGM1,uegaki1,uegaki3,Kamimura:1981oxj,Descouvemont:1987zzb,Tohsaki:2001an,Funaki:2003af,Fedotov:2004nz,Kurokawa:2004ejb,Kurokawa:2005ax,Filikhin:2005nc,Funaki:2005pa,Funaki:2006gt,Arai:2006bt,ohtsubo13,Ishikawa:2014mza,Funaki:2014tda}.
such as the $0^+_2$ state with a cluster gas feature of
weakly interacting three $\alpha$ particles, and
higher $0^+$ and $2^+$ states in the excitation energy $E_x\sim 10 $~MeV region.
Properties and band structure of those cluster
states
are one of the main issues to be clarified.
In spite of the success of $3\alpha$-cluster models in describing many excited
states with cluster structures,
the cluster models fail to describe properties
of low-lying states of $^{12}$C such as
the $2^+_1$ excitation energy and $\beta$-decay transitions from $^{12}$B
because the $\alpha$-cluster breaking is omitted in the models.
Microscopic calculations of $^{12}$C with
the antisymmetrized molecular dynamics (AMD) \cite{KanadaEnyo:1995tb,KanadaEnyo:1995ir,KanadaEn'yo:2012bj} and
Fermionic molecular dynamics \cite{Feldmeier:1994he,Neff:2002nu}
beyond the $3\alpha$-cluster models have been applied to $^{12}$C
and shown that the
$\alpha$-cluster breaking plays an important role
not only in the low-lying states but also in transitions and spectra of cluster states \cite{KanadaEn'yo:1998rf,Kanada-Enyo:2006rjf,Neff:2003ib,Chernykh:2007zz}.
 Furthermore, {\it ab initio} calculations are being developing for structure
study of $^{12}$C \cite{Epelbaum:2012qn,Dreyfuss:2012us,Carlson:2014vla}.

On the experimental side,
the $\alpha$ inelastic scattering has been proved to be
a powerful tool for study of cluster states,
because cluster states can be strongly populated by that process.
For instance,
the $2^+_2$ at 9.84~MeV of $^{12}$C has been recently discovered
with the multipole defomposition analysis (MDA)
in the $^{12}$C$(\alpha,\alpha')$ reaction experiments \cite{Itoh:2011zz,Freer:2012se}.
The $\alpha$ inelastic scattering has been used also for study of
isoscalar monopole and dipole excitations
in a wide energy range.
In the MDA analysis of the $^{12}$C$(\alpha,\alpha')$ reaction,
the significant strengths  have been observed in the low-energy region
below the energy region of the giant resonances
\cite{John:2003ke}, and theoretically described  by the decoupling of
the low-lying cluster modes from the compressive collective vibration modes
of the giant resonances \cite{Kanada-Enyo:2015vwc,Kanada-Enyo:2017fps}.

In order to extract structure information of the excited states,
$\alpha$ elastic and inelastic cross sections have been analyzed
with reaction models
\cite{Itoh:2011zz,John:2003ke,Ohkubo:2004fu,Takashina:2006pc,Khoa:2007as,Takashina:2008zz,Ito:2018opr,Adachi:2018pql}.
To describe these cross sections, many attempts of
the coupled-channel (CC) calculations have been performed
with the optical potentials obtained using
microscopic $3\alpha$-cluster models of $^{12}$C such as the resonating group method
(RGM) \cite{Kamimura:1981oxj} and the $\alpha$ condensation model \cite{Funaki:2006gt}.
However, many of them encountered
the overshooting problem of the $0^+_2$ cross sections, the
so-called {\lq\lq}missing monopole strength''  \cite{Khoa:2007as}.
To circumvent this problem,
phenomenological manipulation of the optical potentials have
been done, for instance, an introduction of state-dependent normalization factors for
the imaginary part of the potentials and the use of density-independent effective
$NN$ interactions instead of the density-dependent ones.

Recently, the $g$-matrix folding model has been developed for study of hadron scattering reactions, and the Melbourne $NN$ interaction \cite{Amos:2000} is found to
successfully describe the nucleon-nucleus and $\alpha$-nucleus scattering cross sections for various nuclei and in a wide range of incident energies.
For the $\alpha$ scattering on $^{12}$C,
the microscopic CC calculation with the Melbourne $g$-matrix interaction has been performed by Minomo and Ogata using the RGM transition density and succeeded to reproduce the $0^+_2$ cross sections as well as the elastic cross sections \cite{Minomo:2016hgc}.
One of the advantages is that
there is no adjustable parameter in the $g$-matrix folding model
because the density- and energy-dependences of the
real and imaginary parts of the effective $NN$ interaction
were determined fundamentally from the $g$-matrix theory.
It turns out that this approach of the $g$-matrix folding model
can be a promising tool to investigate cluster states of general nuclei
by means of the $\alpha$ scattering
if reliable transition densities are provided by structure model calculations.

In this paper, we adopt the $g$-matrix folding model with the Melbourne $NN$ interaction and calculate the cross sections of the $\alpha$ scattering
to the $0^+_{1,2,3}$, $1^-_{1,2}$, $2^+_{1,2}$, $3^-_1$, and
$4^+_{1,2}$ states of $^{12}$C.
The $\alpha$-nucleus CC potentials are derived by folding
the matter and transition densities of $^{12}$C
obtained by a microscopic structure model of the AMD combined
with and without the $3\alpha$-cluster
generator coordinate method (GCM).
The calculated elastic and inelastic cross sections are compared with the observed data at incident energies
of $E_\alpha=130$~MeV, 172.5~MeV, 240~MeV, and
386=MeV\cite{Itoh:2011zz,John:2003ke,Adachi:2018pql,Wiktor1981,Kiss1987}.
The transitions to the $0^+_{2,3}$ and $2^+_2$ states and
also the isoscalar (IS) dipole transitions
to the $1^-_{1,2}$ state are focused.
In the comparison of the present CC calculation with the
DWBA calculation, we discuss the CC effect
to the elastic and inelastic cross sections.
The result obtained with the RGM density is also shown
in comparison with the present result with the AMD density.

The paper is organized as follows.
Sections~\ref{sec:structure-model} and ~\ref{sec:reaction-model} describe the formulations of the
structure and reaction calculations, respectively.
The structure properties of $^{12}$C are shown in Sec.~\ref{sec:structure-results} and
the $\alpha$ scattering cross sections are discussed
in Sec.~\ref{sec:reaction-results}.
Finally, a summary is given in Sec.~\ref{sec:summary}.
The matter and transition densities of $^{12}$C
are shown in appendix~\ref{app:densities}, and definitions of
the transition operators, strengths, and form factors are given
in appendix~\ref{app:operators}.

\section{Structure calculation of $^{12}$C with AMD+VAP with and without $3\alpha$-cluster GCM}\label{sec:structure-model}

The ground and excited states of $^{12}$C are calculated with the variation after projection (VAP) in the AMD framework,
in which the variation is performed for the
spin-parity projected AMD wave function as done in Refs.~\cite{KanadaEn'yo:1998rf,Kanada-Enyo:2006rjf}.
In addition, we combine the AMD+VAP with the $3\alpha$-cluste GCM.
The AMD+VAP and $3\alpha$-cluster wave functions adopted in the present
calculation are the same as those used
in Ref.~\cite{Kanada-Enyo:2015vwc}.
For details of the calculation procedures and wave functions of $^{12}$C,
the reader is referred to those references.

In the AMD method, a basis wave function is given by a Slater determinant,
\begin{equation}
 \Phi_{\rm AMD}({\bvec{Z}}) = \frac{1}{\sqrt{A!}} {\cal{A}} \{
  \varphi_1,\varphi_2,...,\varphi_A \},\label{eq:slater}
\end{equation}
where  ${\cal{A}}$ is the antisymmetrizer, and  $\varphi_i$ is
the $i$th single-particle wave function written by a product of
spatial, spin, and isospin
wave functions,
\begin{eqnarray}
 \varphi_i&=& \phi_{{\bvec{X}}_i}\chi_i\tau_i,\\
 \phi_{{\bvec{X}}_i}({\bvec{r}}_j) & = &  \left(\frac{2\nu}{\pi}\right)^{3/4}
\exp\bigl[-\nu({\bvec{r}}_j-\bvec{X}_i)^2\bigr],
\label{eq:spatial}\\
 \chi_i &=& (\frac{1}{2}+\xi_i)\chi_{\uparrow}
 + (\frac{1}{2}-\xi_i)\chi_{\downarrow}.
\end{eqnarray}
Here $\phi_{{\bvec{X}}_i}$ and $\chi_i$ are the spatial and spin functions, respectively, and
$\tau_i$ is the isospin
function fixed to be proton or neutron. The width parameter $\nu=0.19$ fm$^{-2}$ is used
to minimize the ground state energy of $^{12}$C.
The parameters ${\bvec{Z}}\equiv
\{{\bvec{X}}_1,\ldots, {\bvec{X}}_A,\xi_1,\ldots,\xi_A \}$ indicate
Gaussian centroids and spin orientations, which are treated as variational parameters.
In order to obtain the AMD wave function for the lowest $J^\pi$ state,
the VAP is done as
\begin{eqnarray}
&& \frac{\delta}{\delta{\bvec{Z}}}
\frac{\langle \Phi|H|\Phi\rangle}{\langle \Phi|\Phi\rangle}=0,\\
&&\Phi= P^{J\pi}_{MK}\Phi_{\rm AMD}({\bvec{Z}}),
\end{eqnarray}
where $P^{J\pi}_{MK}$ is the spin-parity projection operator.
For the second and third $J^\pi$ states, the VAP is done for the
component orthogonal to the lower $J^\pi$ states.
One of the advantages of the AMD is that the model is free from
{\it a priori} assumption of clusters because Gaussian centroids
and spin orientations of all single-particle wave functions are independently treated,
but it is able to describe the cluster formation as
well as the cluster breaking.
However, in general, the AMD calculation with a limited number of basis wave functions
is not necessarily enough for a detailed description
of large amplitude inter-cluster motion in developed cluster states.

In order to improve this problem of the AMD,
we explicitly
include the $3\alpha$-cluster wave functions with the GCM. We express
various $3\alpha$-cluster configurations with
the Brink-Bloch cluster wave functions \cite{Brink66}
and superpose them with the AMD+VAP wave functions.
In what follows, we call the AMD+VAP calculation without the $3\alpha$-cluster GCM just the ``AMD'',
and that with the  $3\alpha$-cluster GCM ``AMD+GCM''. In the former calculation, we superpose
23 configurations of the AMD wave functions adopted
in Ref.~\cite{Kanada-Enyo:2006rjf}. In the latter,
150 configurations of the $3\alpha$-cluster are included with the AMD wave functions
as done in Ref.~\cite{Kanada-Enyo:2015vwc}.

As inputs from the structure calculations to the microscopic CC calculation of the $\alpha$ scattering, the matter and transition densities
of $^{12}$C are calculated using the AMD and AMD+GCM
wave functions. The transition strengths and form factors are also calculated and
compared with experimental data determined by
the $\gamma$-decay lifetimes and electron scattering.
The definitions of the densities, strengths, and form factors are given
in Appendixes \ref{app:densities} and \ref{app:operators}.

\section{microscopic coupled-channel calculation with $g$-matrix folding model}\label{sec:reaction-model}

The CC potentials are microscopically derived
by folding the $g$-matrix effective $NN$ interaction with the target
and projectile densities.
We use the Melbourne $g$-matrix interaction \cite{Amos:2000},
which has been successfully used in describing the $\alpha$-nucleus scattering
\cite{Minomo:2016hgc,Egashira:2014zda}.
The $\alpha$-nucleus potential is calculated
with an extended nucleon-nucleus folding (NAF) model.
In this model, first, the nucleon-nucleus CC potentials are obtained by the single folding model using the transition densities of the target nucleus, and then these potentials are folded with the $^4$He one-body density.
For the $^4$He density, we employ the one-range Gaussian density given by Eq.~(24) of Ref.~\cite{Satchler:1979ni}.
The validity of the NAF model for the $\alpha$ elastic scattering
is discussed through the comparison with the so-called target density approximation (TDA) in Ref.~\cite{Egashira:2014zda}. The NAF model is found to well simulate the TDA model and reasonably describe the $\alpha$ elastic scattering on $^{58}$Ni and $^{208}$Pb in a wide range of incident energies of $E_\alpha=20$--200 MeV/u.

It is concluded in Ref.~\cite{Egashira:2014zda} that the TDA model has a clear theoretical foundation in view of the multiple scattering theory and is superior to the conventional frozen density approximation (FDA) in describing the $\alpha$ elastic scattering. Later, the TDA model has successfully been applied to the $^{3}$He elastic scattering \cite{Toyokawa:2015fva} on $^{58}$Ni and $^{208}$Pb, and to the $\alpha$ inelastic scattering on $^{12}$C~\cite{Minomo:2016hgc}. The NAF model adopted in this study will be interpreted as a practical alternative to the TDA model. Nevertheless, there remain some model uncertainties in the reaction calculation, at backward angles in particular.

In the default CC calculation of the elastic and inelastic $\alpha$ scattering,
we adopt the nine states, $0^+_{1,2,3}$, $2^+_{1,2}$, $4^+_{1,2}$, $1^-_{1}$, and $3^-_1$,
of the target $^{12}$C nucleus, with the matter and transition densities
obtained with the AMD and AMD+GCM calculations, which are scaled so as to reproduce
the observed transition strengths to reduce possible ambiguity from the structure calculations.
For the excitation energies of  $^{12}$C, we use the experimental values listed in Table~\ref{tab:radii}.
In the calculation of the $1^-_2$ cross sections with the AMD+GCM, we adopt 13 states
including four states,
$2^+_3$~(12.0 MeV), $2^+_4$~(15.44 MeV), $1^-_2$, and $3^-_2$,
additionally to the above-mentioned nine states.  For the $1^-_2$  and $3^-_2$ states,
which are theoretically predicted in the AMD+GCM calculation,
we choose the excitation energies $E_x=14$ MeV and $E_x=13$ MeV, respectively.

For comparison,
we also perform the CC calculation with the RGM density of $^{12}$C taken from Ref.~\cite{Kamimura:1981oxj},
which have been used in reaction calculations of
the $\alpha$ scattering on $^{12}$C
\cite{Ohkubo:2004fu,Takashina:2006pc,Khoa:2007as,Ito:2018opr,Minomo:2016hgc}.
In the CC calculation with the RGM density,
we adopt five states, the $0^+_{1,2}$, $2^+_{1,2}$, and $3^-_1$, of
$^{12}$C. We do not include the $0^+_{3}$ state of the RGM calculation
because it does not correspond to the physical $0^+_3$ state observed around 10 MeV.

\section{Structure properties of $^{12}$C}\label{sec:structure-results}
In this section, we show structure properties such as radii, transition strengths, and form factors
of the ground and excited states of $^{12}$C obtained with the AMD and AMD+GCM calculations.
For comparison, we also show the RGM result of the $3\alpha$-cluster model from Ref.~\cite{Kamimura:1981oxj}.
Note that, in these structure calculations, there are differences not only in the model wave functions but also in the
effective nuclear interactions. The MV1 central interaction \cite{TOHSAKI} with the Majorana parameter $M=0.62$
and the G3RS \cite{LS1,LS2} spin-orbit interactions with the strength parameters $u_1=-u_2=3000$~MeV
are used in the AMD and AMD+GCM calculations,
whereas the Volkov No.2 central interaction \cite{Volkov:1965zz}  with $M=0.59$ is used in the RGM calculation.

\subsection{Energy spectra and radii of $^{12}$C}

In Table \ref{tab:radii},
excitation energies and root-mean-square (rms) proton radii of the ground and excited states of $^{12}$C
obtained with the structure model calculations of the AMD,  AMD+GCM, and RGM
are listed together with the experimental data.
The AMD and AMD+GCM calculations well reproduce the energy spectra  except for those of
the $4^+_{1,2}$ states, which are somewhat underestimated.
Compared to the RGM, the better reproduction of the $2^+_1$ excitation
energy is obtained in these two calculations because of the $\alpha$-cluster breaking effect.
For the nuclear size of the excited states, three calculations show a trend similar to each other.
Namely, relatively small sizes are obtained for the $2^+_1$ and $4^+_1$ states in the ground band,
whereas much larger sizes than the ground state are obtained for the
developed cluster states such as $0^+_{2,3}$, $1^-_1$, $2^+_2$, $3^-_1$, and $4^+_2$ states.
Quantitatively, the AMD+GCM tends to give slightly larger sizes for the developed
cluster states than the AMD because of the large amplitude cluster motion. Compared with the
two calculations, the RGM shows almost consistent sizes
for  the $0^+_2$ and $2^+_2$ states, but a much smaller size for the $3^-_1$ state than other
two calculations.
In the density profile, one can see qualitatively similar behavior
in the three calculations, but quantitatively, some differences
are found in the central and tail parts of the density.
Comparison of the density between three calculations
is given  in Fig.~\ref{fig:density} of Appendix~\ref{app:densities}.
These differences in the nuclear size and density can be regarded as model ambiguity from structure calculations.

\begin{table}[ht]
\caption{Excitation energies $E_x$ (MeV) and rms proton radii $R_p$ (fm) of $^{12}$C
obtained with the AMD and AMD+GCM calculations.
Theoretical values of the RGM from Ref.~\cite{Kamimura:1981oxj} are also shown.
The experimental energies are taken from Ref.~\cite{Kelley:2017qgh}.
The experimental value of the rms proton radius of the ground state
is deduced from the experimental charge radius measured by the electron scattering \cite{Angeli2013}.
 \label{tab:radii}
}
\begin{center}
\begin{tabular}{lrrrrrrrrccccc}
\hline
\hline
    & \multicolumn{2}{c}{exp}       & \multicolumn{2}{c}{AMD} &   \multicolumn{2}{l}{AMD+GCM} & \multicolumn{2}{c}{RGM} \\
    & $E_x$ & $R_p$   & $E_x$ & $\ \ \ R_p$ & $\ \ \ E_x$ & $R_p$ & $\ \ \ E_x$ & $R_p$ \\
$ 0^+_1 $&  0.0   & 2.33    & 0.0   & 2.53  & 0.0   & 2.54  & 0.0   & 2.40  \\
$ 0^+_2 $&  7.65  &     & 8.1   & 3.27  & 7.3 & 3.62  & 7.74  & 3.47  \\
$ 0^+_3 $&  10.3  &     & 10.7  & 3.98  & 10.0  & 3.92  &   &   \\
$ 1^-_1 $&  10.84   &     & 12.6  & 3.42  & 10.7  & 3.87  &   &   \\
$ 2^+_1 $&  4.44  &     & 4.5   & 2.66  & 4.2 & 2.67  & 2.77  & 2.38  \\
$ 2^+_2 $&  9.87  &     & 10.6  & 3.99  & 9.5 & 4.09  & 9.38  & 3.85  \\
$ 3^-_1 $&  9.64  &     & 10.8  & 3.13  & 9.3 & 3.49  & 8.14  & 2.77  \\
$ 4^+_1 $&  13.3  &     & 10.9  & 2.71  & 10.5  & 2.79  &   &   \\
$ 4^+_2 $&  14.08   &     & 12.6  & 4.16  & 11.6  & 4.22  &   &   \\
\hline
  \hline
\end{tabular}
\end{center}
\end{table}

\subsection{Transition strengths and scaling factors of $^{12}$C}

The transition strengths of $^{12}$C
obtained with the AMD,  AMD+GCM, and RGM calculations
are listed in Table \ref{tab:strengths}
together with the experimental data.
The calculated transition strengths are in reasonable agreement with the experimental data
though the agreement is not perfect.
In order to reduce ambiguity from the structure model calculation,
we introduce the scaling factor
$f_\textrm{tr}=\sqrt{B_\textrm{exp}(E\lambda)/B_\textrm{cal}(E\lambda)}$
(square root of the $B(E\lambda)$ ratio of the experimental value to the theoretical one)
and scale the calculated transition densities as $\rho^\textrm{(tr)}(r) \to f_\textrm{tr} \rho^\textrm{(tr)}(r)$
to fit the experimental $E\lambda$ transition strengths for the use of the $\alpha$ scattering calculation.
The value of $f_\textrm{tr}$  for each transition is shown in Table \ref{tab:strengths}.
For the $1^-_1\to 0^+_1$ transition, 
we determine the scaling factor $f_\textrm{tr}$
by adjusting the calculated charge form factors to the experimental data measure by the electron scattering \cite{Torizuka1969}.
For other transitions with no data of the $E\lambda$ transition strengths,
we set $f_{tr}=1$ and use the calculated transition densities without the scaling, but
the model ambiguity remains.
For instance, for the $0^+_3\to 0^+_1$ transition,  the predicted $B(E0)$ value of the AMD+GCM is
twice as large as that of the AMD.
Also in the transitions of $2^+_2\to 0^+_2$ and $2^+_2\to 0^+_3$,
which are important
for the band assignment of these cluster states near the $3\alpha$ threshold energy,
there are significant differences in the predicted $E2$ strengths between the AMD, AMD+GCM, and RGM calculations.
Even though the transition strengths are adjusted to the
experimental data with the scaling factor, some differences can be seen in detailed behavior
of the calculated transition densities
between the AMD (or AMD+GCM) and RGM.
In Appendix~\ref{app:densities},
we compare the scaled transition densities $f_\textrm{tr}\rho^\textrm{(tr)}(r)$ between three 
calculations. 

\begin{figure*}[!h]
\includegraphics[width=15cm]{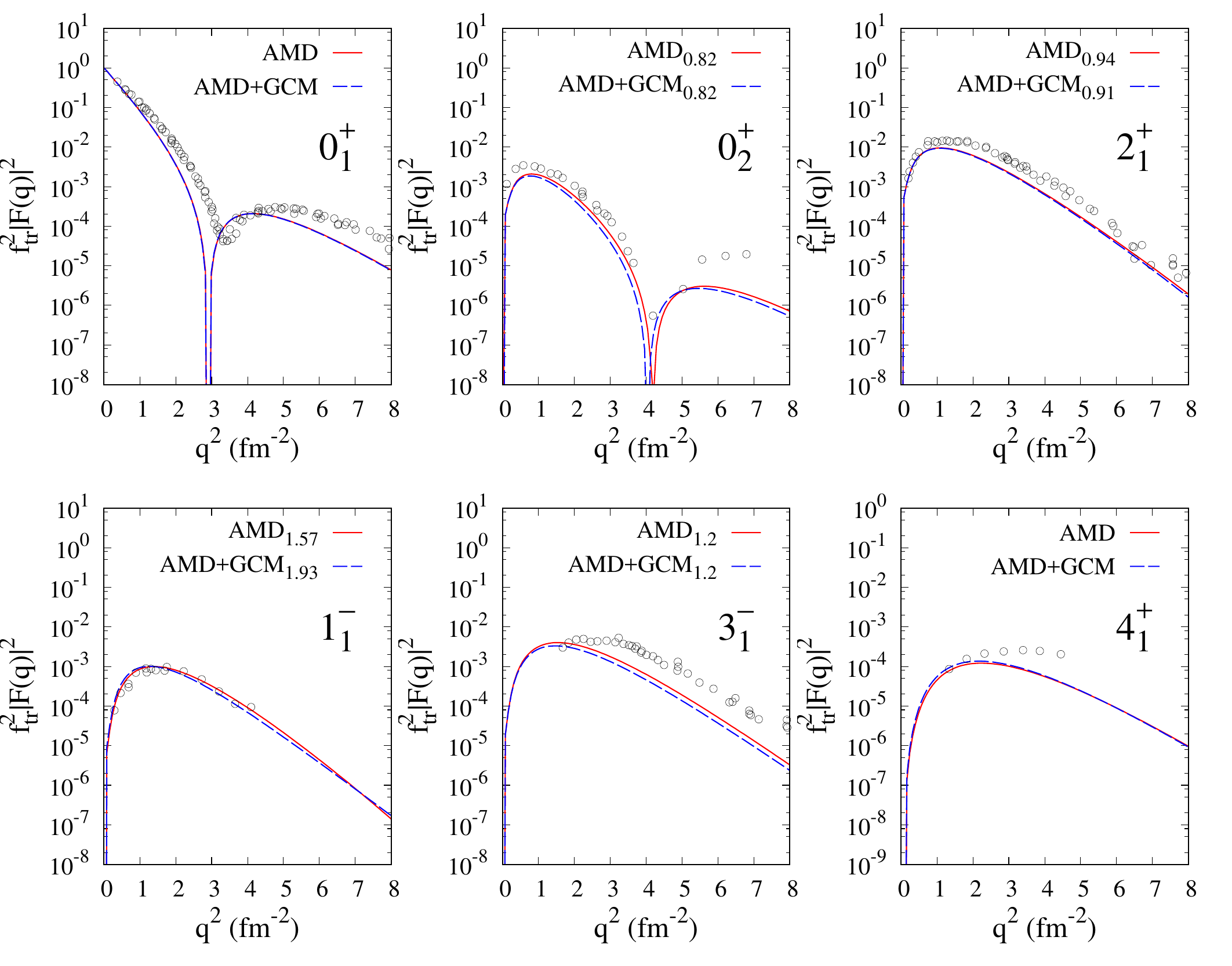}
  \caption{Squared charge form factors of $^{12}$C.
The theoretical values are those obtained with AMD ad AMD+GCM scaled by the factor $f_\textrm{tr}^2$
(labeled by AMD$_{f_\textrm{tr}}$ and AMD+GCM$_{f_\textrm{tr}}$, respectively).
The experimental data are those measured by electron elastic and inelastic scattering on $^{12}$C from
Refs.~\cite{Torizuka1969,Sick:1970ma,Crannel1966,Nakada1971}.
  \label{fig:form}}
\end{figure*}

In Fig.~\ref{fig:form},
the theoretical form factors for electron elastic and inelastic scattering
of the AMD and AMD+GCM are shown compared with the experimental data.
The calculated squared form factors after the scaling
with the factor $f_\textrm{tr}^2$ reasonably agree with the experimental data.

\begin{table*}[ht]
\caption{
The transition strengths $B(E\lambda)$ of $^{12}$C
calculated with the AMD,  AMD+GCM, and RGM.
For the $1^-_1\to 0^+_1$ transition, a quarter
of the isoscalar dipole transition strength $B(\textrm{IS}\lambda)/4$ is shown.
The scaling factors $f_\textrm{tr}=\sqrt{B_\textrm{exp}(E\lambda)/B_\textrm{cal}(E\lambda)}$
determined by the ratio of the experimental strength $B_\textrm{exp}(E\lambda)$
and the calculated strength $B_\textrm{cal}(E\lambda)$ are also shown.
The experimental $B(E\lambda)$ are taken from Ref.~\cite{Kelley:2017qgh}.
$^a$The updated value of $B(E2:2^+_2\to 0^+_1)$ from Ref~\cite{Freer:2014qoa}
by the reanalysis of the data in Ref.~\cite{Zimmerman:2013cxa}.
$^b$The $f_\textrm{tr}$  value for the $1^-_1\to 0^+_1$ transition
is determined by adjusting the charge form factor to the electron scattering data \cite{Torizuka1969}.
The units of the transition strengths are  $e^2~\textrm{fm}^{4}$ for $B(E0)$,
$\textrm{fm}^{6}$ for $B(\textrm{IS1})$, and $e^2~\textrm{fm}^{2\lambda}$ for other  $B(E\lambda)$.
 \label{tab:strengths}
}
\begin{center}
\begin{tabular}{lllllllllccccc}
\hline
\hline
    & \multicolumn{2}{l}{exp} & \multicolumn{2}{l}{AMD} &   \multicolumn{2}{l}{AMD+GCM} & \multicolumn{2}{l}{RGM}   \\
    &     $B(E\lambda)$ & (error) & $B(E\lambda)$ & $f_\textrm{tr}$ & $B(E\lambda)$ & $f_\textrm{tr}$ & $B(E\lambda)$ & $f_\textrm{tr}$ \\
$ E2:2^+_1\to 0^+_1 $&  7.59  &$( 0.42  )$  & 8.53  & 0.94  & 9.09  & 0.91  & 9.31  & 0.90  \\
$ E0:0^+_2\to 0^+_1 $&  29.2  &$( 0.2   )$  & 43.5  & 0.82  & 43.3  & 0.82  & 43.8  & 0.82  \\
$ E2:0^+_2\to 2^+_1 $&  13.5  &$( 1.4   )$  & 25.1  & 0.73  & 24.1  & 0.75  & 5.6   & 1.56  \\
$ E2:2^+_2\to 0^+_1 $&  1.57$^a$  &$( 0.13  )$  & 0.39  & 1.99  & 0.49  & 1.93  & 2.48  & 0.80  \\
$ E2:3^-_1\to 1^-_1 $&          &&  40.7  & 1 & 79.0  & 1 &   &   \\
$ E0:0^+_3\to 0^+_1 $&          &&  5.2   & 1 & 10.0  & 1 &   &   \\
$ \textrm{IS1}:1^-_1\to 0^+_1 $&          &&  2.6   & 1.57$^b$  & 2.4   & 1.93$^b$  & 5.7 & 1 \\
$ \textrm{IS1}:1^-_2\to 0^+_1 $&          &&    &   & 1.5   & 1 &   &   \\
$ E3:3^-_1\to 0^+_1 $&  103   &$( 17  )$  & 71  & 1.20  & 71  & 1.20  & 125   & 0.91  \\
$ E4:4^+_1\to 0^+_1 $&          &&  733   & 1 & 995   & 1 & 655 & 1 \\
$ E3:3^-_1\to 0^+_2 $&          &&  428   & 1 & 1210  & 1 & 228   & 1 \\
$ E2:2^+_2\to 0^+_2 $&          &&  102   & 1 & 182   & 1 & 212   & 1 \\
$ E2:2^+_2\to 0^+_3 $&          &&  309   & 1 & 223   & 1 &   &   \\
\hline
\hline
\end{tabular}
\end{center}
\end{table*}

\section{$\alpha$ scattering cross sections}\label{sec:reaction-results}

The cross sections of the $^{12}$C$(\alpha,\alpha')$ reaction
at incident energies of $E_\alpha=130$ MeV, 172.5 MeV, 240 MeV, and 386 MeV
are calculated by the CC calculation with the $g$-matrix folding
potentials using the the theoretical transition densities scaled by the factor $f_\textrm{tr}$.
The cross sections obtained with the AMD, AMD+GCM, and RGM densities
are discussed in comparison with experimental data.
The cross sections obtained by the
DWBA calculation are also shown to discuss the CC effect.

\subsection{Cross sections with the AMD and AMD+GCM}

\begin{figure*}[!h]
\includegraphics[width=15cm]{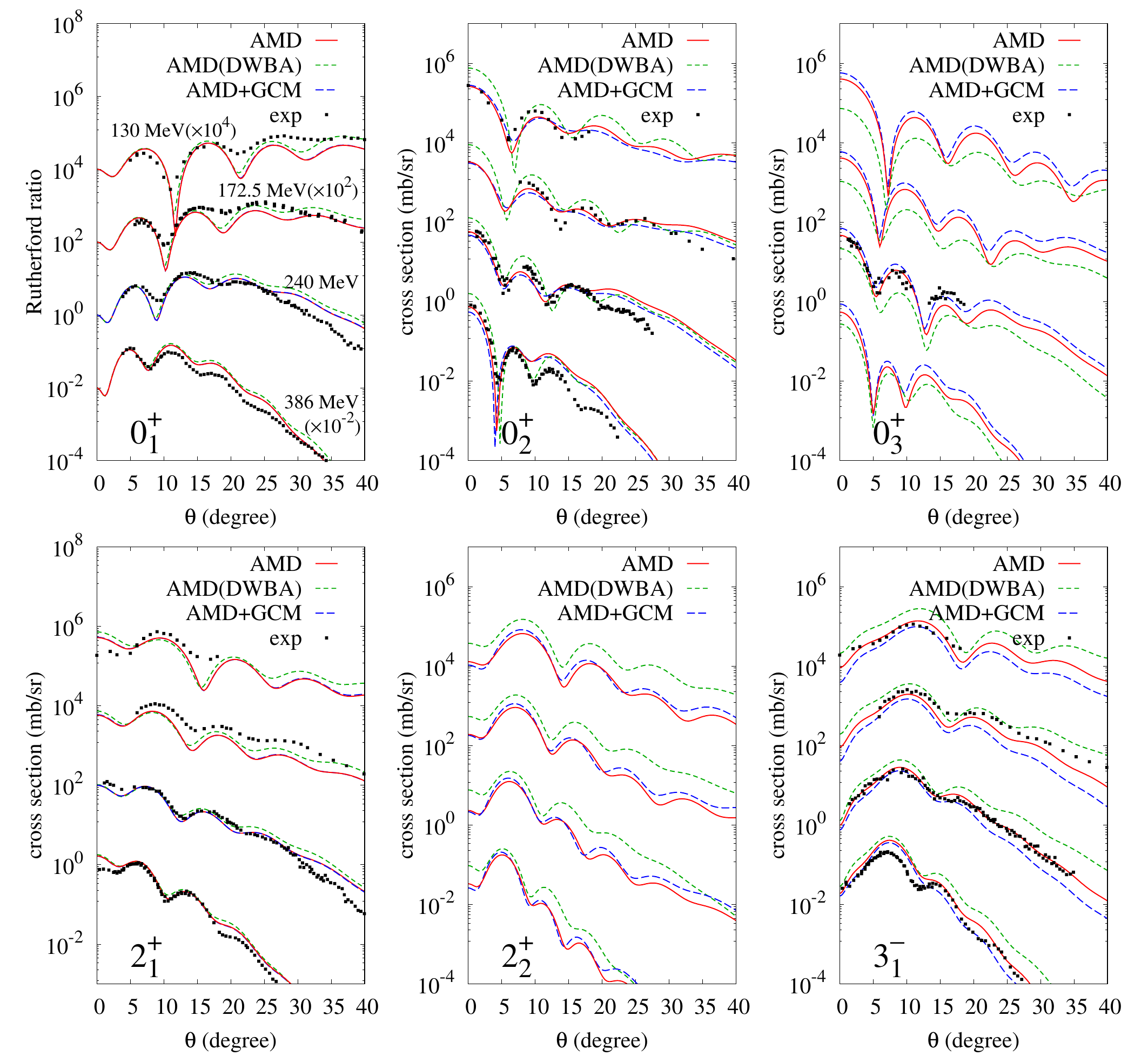}
 \caption{
$\alpha$ scattering cross sections
on $^{12}$C  at $E_\alpha=130$ MeV ($\times 10^4$), 172.5 MeV  ($\times 10^2$), 240 MeV, and 386 MeV  ($\times 10^{-2}$),
The differential cross sections of the $0^+_{1,2,3}$, $2^+_{1,2}$, and $3^-_1$
states
obtained by the CC calculation with the AMD and AMD+GCM
transition densities are shown.
The cross sections obtained by the  DWBA calculation
with the AMD transition densities are also shown
for comparison.
The experimental data are taken
from Refs.~\cite{John:2003ke,Itoh:2011zz,Adachi:2018pql,Kiss1987,Wiktor1981}.
References for those data
are summarized in Table \ref{tab:cross-data}.
\label{fig:cross-gcm1}}
\end{figure*}

\begin{figure*}[!h]
\includegraphics[width=15cm]{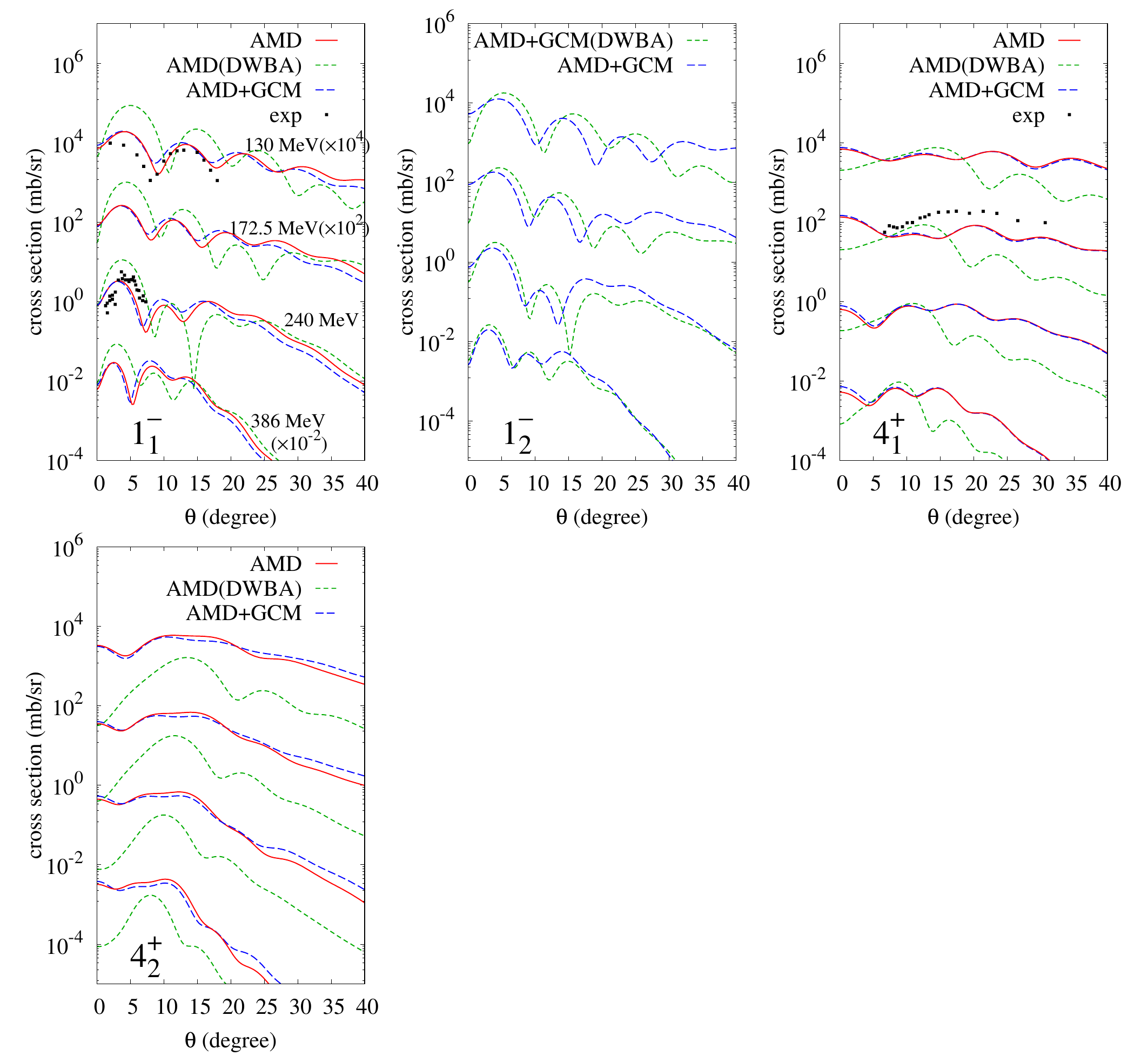}
  \caption{Same as Fig.~\ref{fig:cross-gcm1}, but for the $1^-_{1,2}$ and $4^+_{1,2}$ states.
  \label{fig:cross-gcm2}}
\end{figure*}

In Figs.~\ref{fig:cross-gcm1} and \ref{fig:cross-gcm2},
the calculated cross sections with the AMD (solid lines) and AMD+GCM (dashed lines)
are shown together with experimental data.
The cross sections obtained by the DWBA calculation with the AMD
are also shown by the dotted lines.

The obtained cross sections are qualitatively similar to each other
between the AMD and AMD+GCM.
These calculations reasonably reproduce the cross sections for The elastic scattering and the inelastic scattering to the $0^+_{2,3}$, $2^+_1$, $1^-_1$, and $3^+_1$ states.
It should be stressed again that the present microscopic CC calculation with the $g$-matrix folding potentials
contains no adjustable parameter except for the scaling factor to fit the data of the electric
transition strengths, $B(E\lambda)$.
It indicates the applicability of the present model for the $\alpha$ scattering on $^{12}$C
in this energy region of $E_\alpha=130$--400~MeV.

In the $0^+_2$ cross sections, one can see that
the amplitudes of the
first and second peaks are reproduced well, and there is no overshooting problem of the 
$0^+$ cross sections for this state
as in Ref.~\cite{Minomo:2016hgc}.
In the $0^+_3$ inelastic cross sections, two calculations of the AMD and AMD+GCM show
a slight difference in the absolute amplitude: the AMD+GCM shows about 1.5 times larger cross sections
than the AMD because of the larger $E0$ strength for the direct transition $0^+_1\to 0^+_3$,
but both reasonably describe the experimental cross sections taken at $E_\alpha=240$ MeV  \cite{John:2003ke}.
It should be remarked that the data corresponding to the broad resonance around 10.3 MeV,
and it can contain two $0^+$ states as reported recently \cite{Itoh:2011zz}.

For the $2^+_1$ cross sections, there is no difference between the AMD and AMD+GCM. Both reproduce
the  cross sections with comparable quality to the elastic scattering.
As for the $3^-_1$ cross sections, the AMD and AMD+GCM show a quantitative difference in the absolute amplitude
even though the $E3$ transition strength is adjusted to the experimental value in both cases.
The AMD+GCM gives somewhat smaller cross sections than the AMD. A possible reason for this
is the larger radius of the $3^+_1$ state in the AMD+GCM, which may cause stronger absorption
than in the AMD.

For the $1^-_1$ cross sections, the AMD and AMD+GCM results are consistent with each other, and
both are in reasonable agreement with the experimental data at $E_\alpha=240$~MeV. Because
the scaled transition density can reproduce both the electric scattering and $\alpha$ scattering data, we can
estimate the IS dipole transition strength as $B(\textrm{IS1};1^-_1\to 0^+_1)/4=6$--9~fm$^6$.

\begin{figure}[!h]
\includegraphics[width=10cm]{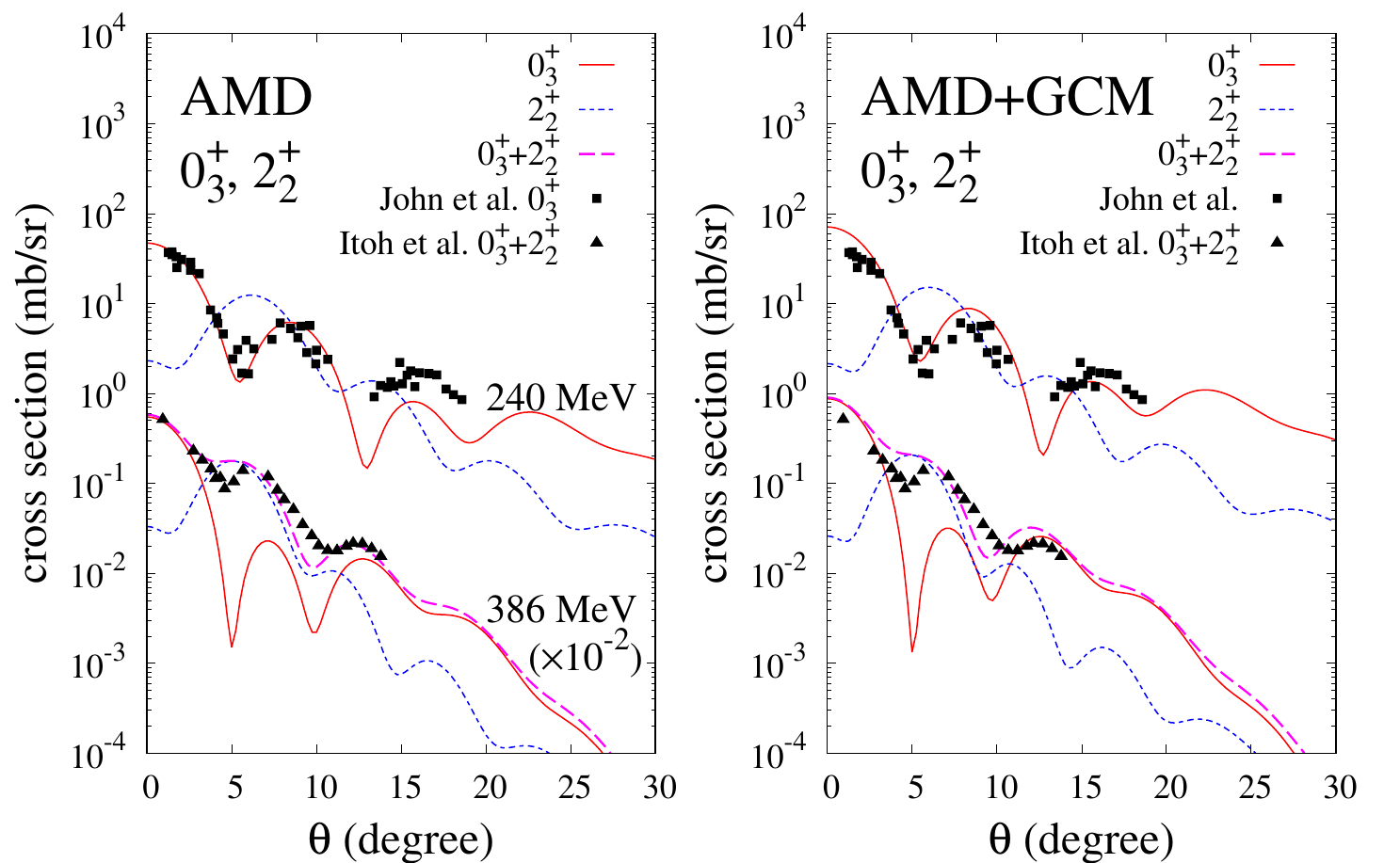}
  \caption{
$2^+_2$ and $0^+_3$ cross sections at $E_\alpha=240$ MeV and  386 MeV calculated with the AMD and AMD+GCM.
The incoherent sum of the $2^+_2$ and $0^+_3$ cross sections at 386 MeV is compared with the
experimental sum of the $2^+_2(9.84)$~MeV and $0^+_3$(9.93~MeV) taken from
Ref.~\cite{Itoh:2011zz}. The experimental $0^+_3$ cross sections at 240 MeV  are taken  from Ref.~\cite{John:2003ke}.
  \label{fig:cross-s0-0_3}}
\end{figure}

The $2^+_2$ state is the newly discovered state by
$\alpha$ inelastic scattering and $\beta$-decay experiments \cite{Freer:2014qoa,Itoh:2011zz,Freer:2012se}.
The predicted cross sections of the $2^+_2$ state are
much smaller than the $2^+_1$ state consistently with the
weak $E2$ transition from the $0^+_1$, a small $B(E2;2^+_2\to 0^+_1)$, because this state is the cluster
state and has the strong $E2$ transitions not to the ground state but to the $0^+_2$ and $0^+_3$ states.
In Fig.~\ref{fig:cross-s0-0_3}, we compare the incoherent sum of the
$2^+_2$ and $0^+_3$ cross sections at 386 MeV compared
with the experimental sum of the $2^+_2(9.84)$~MeV and $0^+_3$(9.93~MeV)
reported in Ref.~\cite{Itoh:2011zz}. The $2^+_2$ and $0^+_3$ cross sections at 240 MeV are also shown together with the
experimental  $0^+_3$ cross sections.
In the calculation, the $0^+_3$ and $2^+_2$ cross sections  describe respectively
the first and second peaks, and both contribute to the third peak of the summed cross sections.
This result is similar to the experimental MDA analysis \cite{Itoh:2011zz} and
the theoretical calculation of Ref.~\cite{Ito:2018opr},
where the optical potentials have been phenomenologically tuned to reproduce the experimental cross sections.
In the reproduction of the experimental data, the AMD result seems to be favored rather than the AMD+GCM, though
quality of the reproduction is not satisfactory to conclude it.

For the $1^-_2$, and $4^+_2$ states, there are no available data
and the calculated cross sections are theoretical predictions.
As discussed in Ref.~\cite{Kanada-Enyo:2017fps}, the predicted $1^-_2$ is a toroidal dipole state
and contributes to the isoscalar dipole strengths in the low-energy region below the giant dipole resonance.
In the $\alpha$ scattering experiment at 240 MeV \cite{John:2003ke},
the significant isoscalar dipole strength around 15 MeV has been observed in the
MDA, and it is a candidate for the predicted toroidal state of the $1^-_2$.


\begin{table}[ht]
\caption{References for experimental differential cross sections of the
$\alpha$ scattering on $^{12}$C at incident energies of $E_\alpha=130$ MeV, 172.5 MeV, 240 MeV, and 386 MeV.
$^a$The excitation energy of the $0^+_3$ state (the broad resonance around 10 MeV)
is 10.3 MeV in Ref.~\cite{John:2003ke} and 9.93 MeV in Ref.~\cite{Itoh:2011zz}.
$^b$The sum of the cross sections of the $2^+_2$(9.84 MeV) and $0^+_3$(9.93 MeV).
 \label{tab:cross-data}}
\begin{center}
\begin{tabular}{cccccccccc}
\hline
$J^\pi_f$ ($E_x$) & 130 MeV & 172 MeV & 240 MeV   & 386 \\
$0^+_1$(0.00) &\cite{Adachi:2018pql} & \cite{Kiss1987},\cite{Wiktor1981} &  \cite{John:2003ke}& \cite{Itoh:2011zz} \\
$2^+_1$(2.44) &\cite{Adachi:2018pql}  &\cite{Kiss1987} &  \cite{John:2003ke} & \cite{Itoh:2011zz} \\
$0^+_2$(7.65) & \cite{Adachi:2018pql}  & \cite{Kiss1987} & \cite{John:2003ke} &
\cite{Itoh:2011zz},\cite{Adachi:2018pql}\\
$0^+_3$(10.3$^a$) &  & & \cite{John:2003ke} & \cite{Itoh:2011zz}$^{b}$  \\
$2^+_2$(9.84) &  & & &\cite{Itoh:2011zz}$^b$ \\
$3^-_1$(9.64) & \cite{Adachi:2018pql}  &\cite{Kiss1987}  &  \cite{John:2003ke} &\cite{Itoh:2011zz},\cite{Adachi:2018pql}\\
$1^-_1(10.84)$ & \cite{Adachi:2018pql} & & \cite{John:2003ke}  & \\
$4^+_1$(14.0) &  & \cite{Kiss1987} & & \\
\hline
\end{tabular}
\end{center}
\end{table}

\subsection{Coupled-channel effects}
Let us discuss the CC effect in comparison with the
DWBA calculation shown in Figs.~\ref{fig:cross-gcm1} and \ref{fig:cross-gcm2}.

For the $0^+_1$ and $2^+_1$ cross sections,
the results are almost consistent between the DWBA and CC calculations, and
only a slight difference can be seen at large scattering angles.
For other states, the CC effect is significant, in particular, at low incident energies, and still remains
even at $E_\alpha=386$ MeV.
In the $0^+_2$, $2^+_2$, and  $3^-_1$ cross sections, the absolute amplitudes are reduced
by the CC effect. Compared with the DWBA calculation,
the peak positions are almost unchanged but dips are somewhat smeared
in the CC calculation for the $0^+_2$ and $2^+_2$.
The CC effect on the  $0^+_2$ cross sections is dominantly contributed by
the $\lambda=2$ transitions with the $2^+_1$ and $2^+_2$ and $\lambda=3$ transition with the $3^-_1$.
The CC effect on the $2^+_2$ cross sections turn out to be through the $\lambda=2$ transition with the $0^+_2$ and
the $\lambda=3$ transition with the $3^-_1$,

For the  $0^+_3$ cross sections, the CC effect gives an opposite contribution, namely, it enhances the cross sections.
Consequently, the calculated $0^+_3$ cross sections are the same order as the  $0^+_2$
cross sections even though the monopole transition strength to the $0^+_3$ is about one order
smaller than the strength to the $0^+_2$.
This result indicates that the $0^+$ cross sections do not scale
with the monopole transition strengths contrary to the naive expectation
of the linear scaling, which is often assumed in the experimental
determination of isoscalar transition strengths with
the DWBA analysis of the $\alpha$ inelastic scattering.

Further significant CC effects are found in the $1^-_1$, $4^+_1$, and $4^+_2$ cross sections.
For these states, not only the absolute amplitude  but also the diffraction pattern of the cross sections are affected.
For the $1^-_1$ cross sections, the absolute values are reduced and the first and second
peak positions are shifted to the forward angle by the CC effect, which is
essential to describe the experimental cross sections at 130 MeV.
The dominant contribution to the $1^-_1$ cross sections is the coupling with the $3^-_1$ state
through the strong $\lambda=2$ transition.
Compared with the $1^-_1$ case,
the CC effect in the $1^-_2$ cross sections is not so large.
The present calculation predicts almost the same amplitude of the $1^-_2$ cross sections as the $1^-_1$ cross sections
even though the isoscalar dipole transition strength is weaker in the $1^-_2 \to 0^+_1$ than in the $1^-_1 \to 0^+_1$
as shown in Table \ref{tab:strengths}.

For the $4^+_1$ and $4^+_2$  states, the cross sections  are strongly influenced by the channel coupling.
For the $4^+_1$ cross sections,  the present CC calculation reproduces the absolute amplitude but
does not describe the diffraction pattern of the experimental cross sections.

\subsection{Cross sections with the RGM}

\begin{figure*}[!h]
\includegraphics[width=15cm]{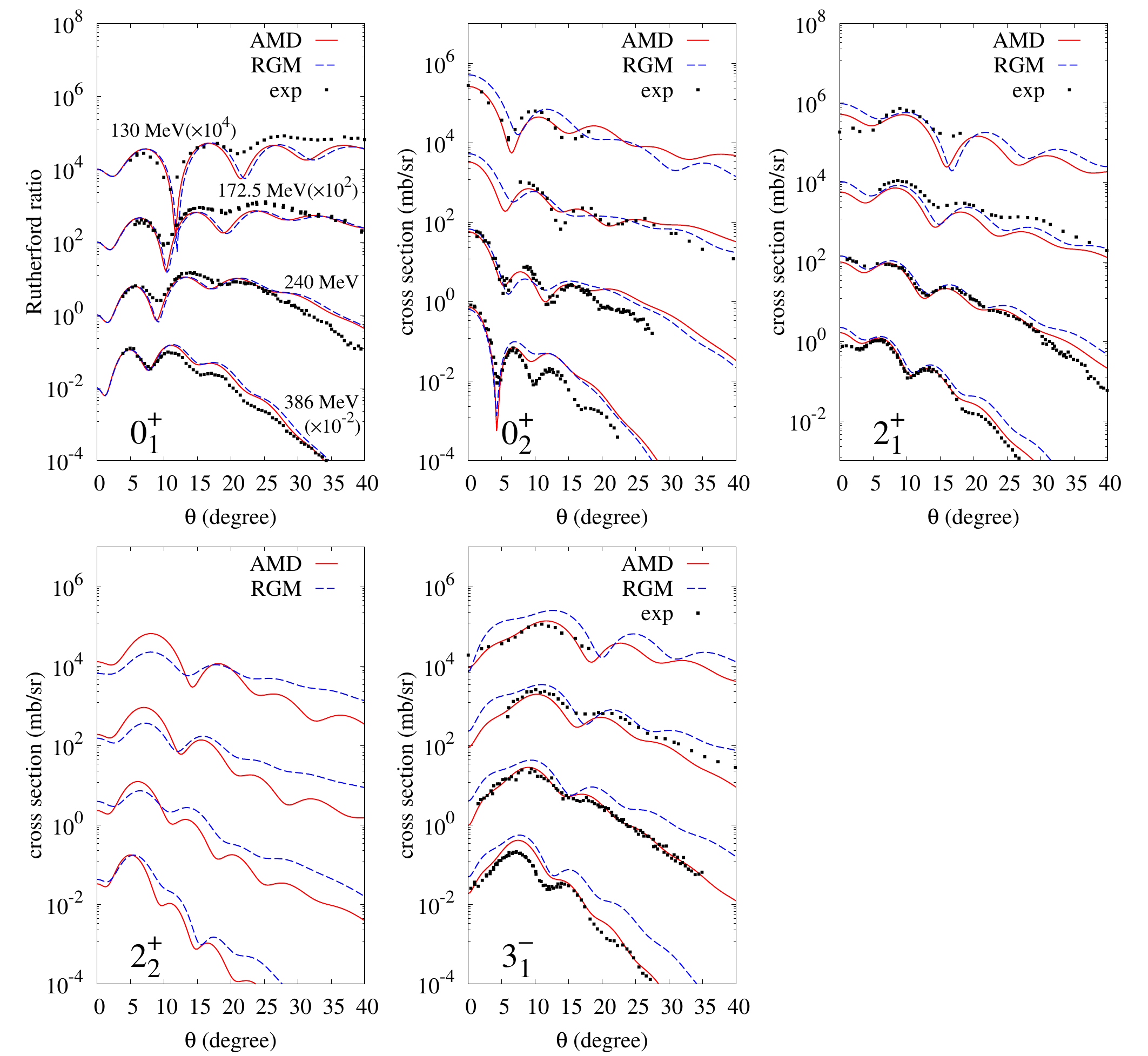}
  \caption{
$\alpha$ scattering cross sections
on $^{12}$C  at $E_\alpha=130$ MeV ($\times 10^4$), 172.5 MeV  ($\times 10^2$), 240 MeV, and 386 MeV  ($\times 10^{-2}$),
obtained by the CC calculation with the RGM densities compared with the AMD result.
The calculated differential cross sections of the $0^+_{1,2}$, $2^+_{1,2}$, and $3^-_1$
states are shown.
The experimental data
from Refs.\cite{John:2003ke,Itoh:2011zz,Adachi:2018pql,Kiss1987,Wiktor1981} are also shown.
  \label{fig:cross-rgm}}
\end{figure*}

Figure \ref{fig:cross-rgm} shows the cross sections obtained with the RGM together with the
AMD result as well as the experimental data.
Some differences can be seen in the inelastic cross sections between the RGM and AMD.
The RGM shows larger cross sections for the $3^-_1$ than the AMD, and tends to overestimate the experimental data.
The absorption may be too weak in the RGM because of the smaller radius of the $3^-_1$ state
than the AMD result.
For the $0^+_2$ cross sections,
the peak and dip structures are smeared by the stronger CC effect in the RGM result,
and the reproduction of the experimental data becomes
somewhat worse than the AMD.
Also in the $2^+_2$ cross sections, the strong CC effect smears the diffraction pattering in the RGM result.

\section{Summary}\label{sec:summary}

The $\alpha$ elastic and inelastic scattering on $^{12}$C was investigated by the
microscopic CC calculation with the $g$-matrix folding model.
The $\alpha$-nucleus CC potentials are derived by folding the Melbourne $g$-matrix $NN$ interaction
with the transition densities calculated with the microscopic structure models of
the AMD and AMD+GCM.

The present calculation reasonably reproduces the differential cross sections of the $\alpha$  scattering
at incident energies of $E_\alpha=130$ MeV, 172.5 MeV, 240 MeV, and 386 MeV
with no adjustable parameter
except for the scaling factor to fit the data of the electric
transition strengths, $B(E\lambda)$.
The calculation successfully describes the absolute amplitude of the $0^+_2$ cross sections
and does not encounter the overshooting problem of the $0^+$ cross sections, 
the so-called missing monopole strength. This result is consistent with the preceding work by Minomo and Ogata \cite{Minomo:2016hgc}
using the RGM transition densities.
Moreover, the present calculation reproduces the $0^+_3$ cross sections and also describes
the sum of the $0^+_3$ and $2^+_2$ cross sections.
In comparison with the DWBA calculation, the CC effect on the inelastic scattering
cross sections except for the $2^+_1$ cross sections
is found to be significant, in particular, at low incident energies, and
still remains even at $E_\alpha=386$ MeV.

It was found that the absolute values of the inelastic cross sections do not
necessarily scale linearly with the transition strength, because
it is sensitively influenced by the coupling with other channels and also by the radius of the excited state.
This may be a characteristic aspect of the $\alpha$ scattering on $^{12}$C, in which
cluster states near the threshold energy have larger radii than the states  in the ground band states
and there exist strong transitions between each other.
It indicates that reliable microscopic calculation of $\alpha$ scattering is needed to
extract quantitative information on the transition strengths from the $\alpha$ inelastic scattering.
It should be remarked that such calculation may reveal also properties of the coupling between excited states that
cannot be studied if the DWBA picture holds. The $\alpha$ inelastic cross sections contain rich information on the excited
states of $^{12}$C through the CC effect.
The present model has been proved to be applicable to the $\alpha$ elastic and inelastic scattering for cluster states and
can be a powerful tool for investigation of not only the
isoscalar monopole and dipole transitions but also transitions between excited states for general stable and unstable nuclei.

Nevertheless, there still remain problems in an accurate reproduction of the cross sections.
There is no ambiguity for the known transitions because
the theoretical transition densities are scaled to fit existing data of the transition strengths. However,
 for unknown transitions, in particular, transitions between excited states, model
ambiguity remains in the structure calculations. Another unknown factor is
the nuclear size of the excited states.
Further reliable structure calculations are needed to reduce the ambiguity from these factors.
Also in the reaction part, further improvements can be considered.
For example, treatments of the density dependences of the $g$-matrix effective interactions and
a possible contribution of the three-nucleon force effect should be tested more carefully for
better reproduction of the scattering cross sections, those at backward angles in particular.

\begin{acknowledgments}
The computational calculations of this work were performed by using the
supercomputer in the Yukawa Institute for theoretical physics, Kyoto University. This work was supported
in part
by Grants-in-Aid of the Japan Society for the Promotion of Science (Grant Nos. JP26400270, JP18K03617, and JP16K05352).
\end{acknowledgments}

\appendix
\section{Matter and transition densities}\label{app:densities}

\begin{figure*}[!h]
\includegraphics[width=15 cm]{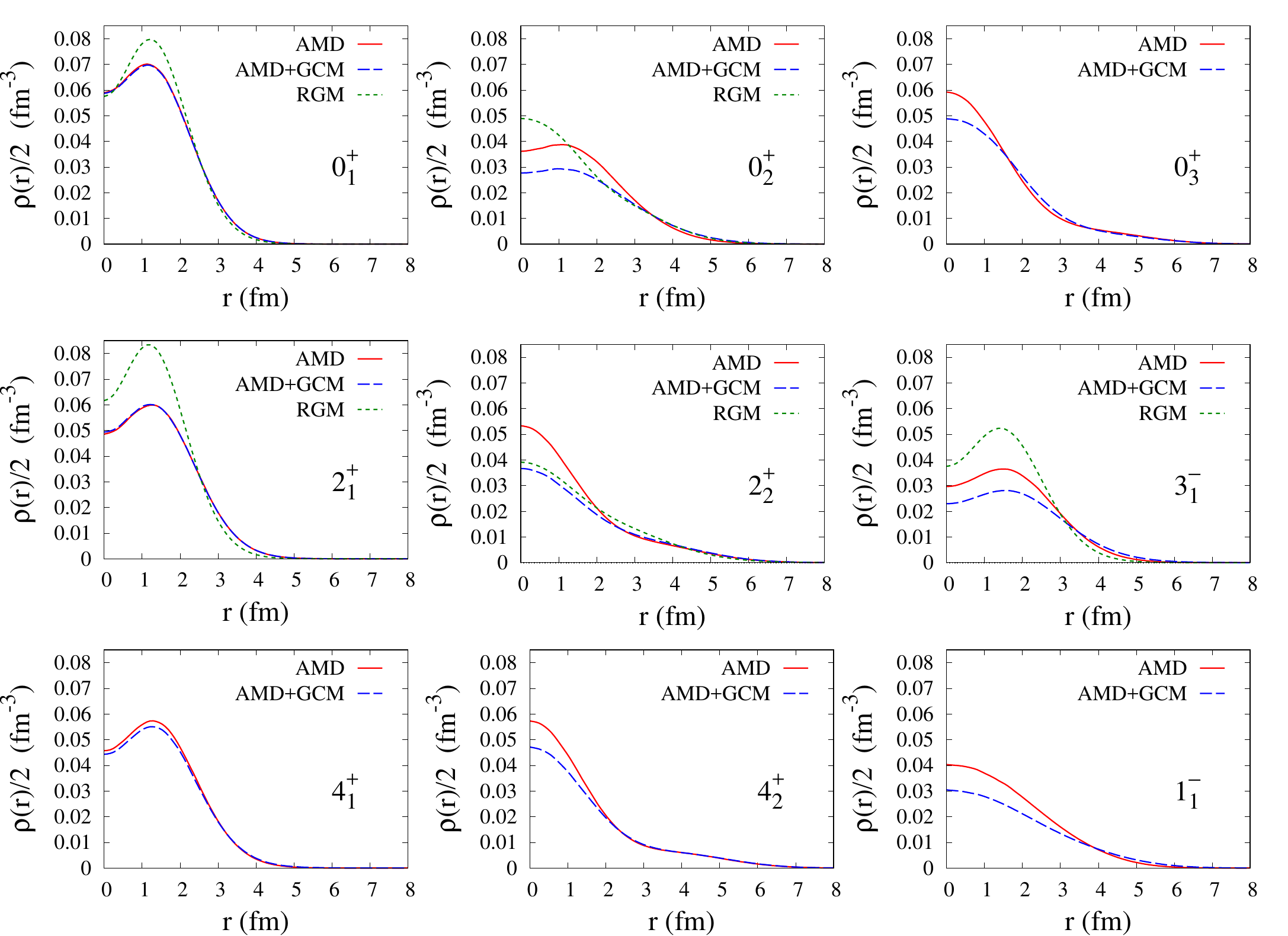}
  \caption{proton densities $\rho_p(r)=\rho(r)/2$.of the 
AMD, AMD+GCM, and RGM calculations.
  \label{fig:density}}
\end{figure*}

\begin{figure*}[!h]
\includegraphics[width=15cm]{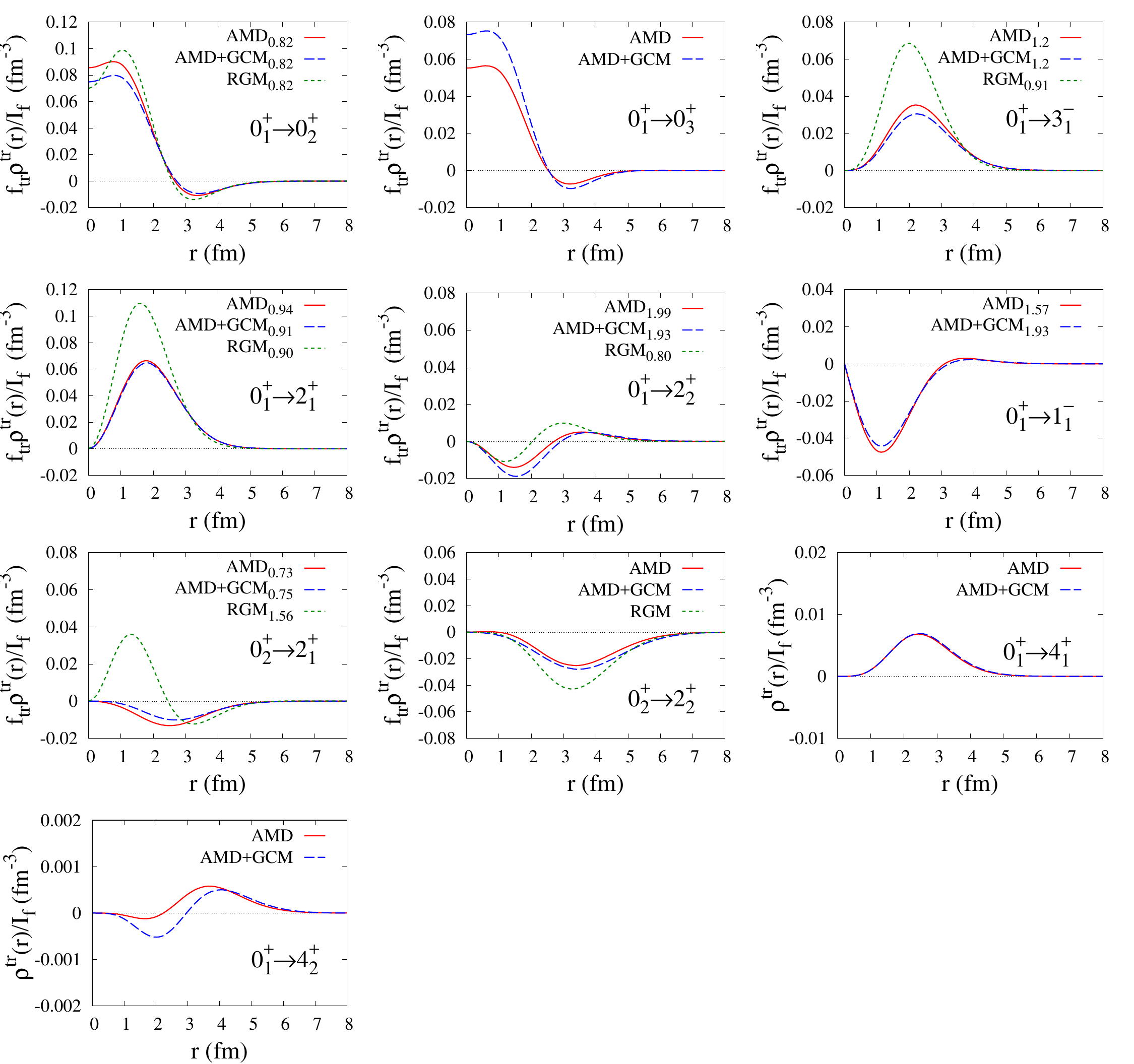}
  \caption{transition densities $\rho^\textrm{tr}(r)$ of the 
AMD, AMD+GCM, and RGM. calculations.
The calculated densities scaled with a factor $I_f\equiv \sqrt{2J_f+1}$. are plotted.
  \label{fig:trans}}
\end{figure*}

The density operator of nuclear matter is
\begin{eqnarray}
\rho(\bvec{r})&=& \sum_k  \delta(\bvec{r}-\bvec{r}_k).
\end{eqnarray}
The transition density for the transition $|i \rangle \to |f\rangle$ is given as
$\rho^\textrm{(tr)}_{i \to f}(\bvec{r})\equiv \langle f |\rho(\bvec{r}) |i \rangle$, and
its $\lambda$th moment is obtained
from the multipole decomposition,
\begin{align}
\rho^\textrm{(tr)}_{i\to f}(\bvec{r})&=\frac{1}{\sqrt{2J_f+1}}\sum_\lambda \rho^\textrm{(tr)}_{\lambda;i\to f}(r)
\nonumber\\
&\times
\sum_\mu Y^*_{\lambda\mu}(\hat{\bvec{r}})(J_i M_i \lambda \mu|J_f M_f),
\end{align}
where $J_i$ and $M_i$  ($J_f$ and $M_f$) are the spin quantum numbers of the initial $|i\rangle$ (final $|f\rangle$) state.
It should be remarked that the transition density $\rho^\textrm{(tr)}_{\lambda;i\to f}(r)$ defined here
is related to the transition density ${\rho}^\textrm{(tr:K)}_{\lambda;i\to f}(r)$
used by Kamimura in Ref.~\cite{Kamimura:1981oxj} as
\begin{align}
{\rho}^\textrm{(tr:K)}_{\lambda;i\to f}(r)=\frac{1}{\sqrt{2J_f+1}}\rho^\textrm{(tr)}_{\lambda;i\to f}(r).
\end{align}
The matter density $\rho(r)$ of the state $|i\rangle$
is related to the diagonal component of the $\lambda=0$ transition density as
\begin{align}
 \rho(r)=\frac{1}{\sqrt{4\pi}} \rho^\textrm{(tr)}_{0;i\to i}(r).
\end{align}The volume integral of  the matter density equals to the mass number $A$ as
\begin{align}
A=\int 4\pi r^2 \rho(r) dr.
\end{align}
The matter and transition densities obtained with the AMD, AMD+GCM, and RGM calculations are
shown in Figs.~\ref{fig:density} and \ref{fig:trans}, respectively.

\section{Definitions of transition operators, strengths, and form factors}
\label{app:operators}
For the rank $\lambda\ne 0,1$,
the isosalar transition operator is give as
\begin{align}
M_{\textrm{IS}\lambda}(\mu)\equiv&\int d\bvec{r} \rho(\bvec{r}) r^{\lambda} Y_{\lambda\mu} (\hat{\bvec{r}}),
\end{align}
and the matrix element is related to
the transition density as
\begin{align}
\langle f || M_ {\textrm{IS}\lambda} ||i \rangle = \int dr r^2 r^{\lambda}
\rho^\textrm{(tr)}_{\lambda;i\to f}(r).
\end{align}
In the preset calculation, the electric transitions are calculated by assuming the mirror symmetry
because the symmetry breaking in the initial and final states are negligibly small.
The $E\lambda$ transition strength is given as
\begin{align}
B(E\lambda)=\frac{e^2}{4}\frac{1}{2J_i+1} \left| \langle f||M_ {\textrm{IS}\lambda}||i \rangle \right |^2,
\end{align}
where the factor of $\frac{1}{4}$ comes from the mirror symmetry assumption.
For the $\lambda=0$ case, the  $E0$ transition operator, matrix elements, and strengths are given as
\begin{align}
M_\textrm{IS0}&\equiv\int d\bvec{r} \rho(\bvec{r}) r^2,\\
\langle f||M_ {\textrm{IS}0} ||i \rangle &= \sqrt{4\pi} \int dr r^2 r^{\lambda+2}
\rho^\textrm{(tr)}_{\lambda;i\to f}(\bvec{r}),\\
B(E0)&=\frac{e^2}{4}\frac{1}{2J_i+1} \left| \langle f||M_ {\textrm{IS}0}||i \rangle \right |^2.
\end{align}
The $\lambda$th multipole component of the so-called longitudinal form factor is related to the Fourier-Bessel transform of the transition
charge density
$\rho^\textrm{ch}_{\lambda;i\to f}(r)$ by
\begin{align}
F(q)=
\frac{\sqrt{4\pi}}{Z}\frac{1}{\sqrt{2J_i+1}} \int dr r^2 j_\lambda(qr)  \rho^\textrm{ch}_{\lambda;i\to f}(r),
\end{align}
where  $\rho^\textrm{ch}_{\lambda;i\to f}(r)$ is calculated by taking into account the proton charge radius.

\end{document}